\documentclass[a4paper,11pt]{article}
\pdfoutput=1   

\usepackage{subfigure}
\usepackage{amsmath}
\usepackage{rotating}
\usepackage{mathtools}
\usepackage{graphicx}
\usepackage{textpos}
\usepackage{float}
\usepackage{color}

\definecolor{orange}{rgb}{1,0.7,0}
\definecolor{lightpurple}{rgb}{1,0.7,1}

\DeclareMathOperator{\Tr}{T\ \! \! r}

\pdfoutput=1 

\usepackage{jheppub} 

\usepackage[T1]{fontenc} 

\def\ltap{\;\raisebox{-.5ex}{\rlap{$\sim$}} \raisebox{.5ex}{$<$}\;}
\def\gtap{\;\raisebox{-.5ex}{\rlap{$\sim$}} \raisebox{.5ex}{$>$}\;}

\title{The Charmonium Potential at Non-Zero Temperature}

\author[a]{Chris Allton,}
\author[a,b]{Wynne Evans,}
\author[c]{Pietro Giudice,}
\author[d]{and Jon-Ivar Skullerud}

\affiliation[a]{Swansea University, Swansea, United Kingdom}
\affiliation[b]{Bern University, Bern, Switzerland}
\affiliation[c]{Universit\"at M\"unster, M\"unster, Germany}
\affiliation[d]{University of Maynooth, Maynooth, Ireland}

\emailAdd{pyevans@swansea.ac.uk}
\emailAdd{c.allton@swansea.ac.uk}
\emailAdd{p.giudice@uni-muenster.de}
\emailAdd{jonivar@thphys.nuim.ie}

\abstract{The potential between charm and anti-charm quarks is
  calculated non-perturb-atively using physical, rather than static
  quarks at temperatures on both sides of the deconfinement transition
  $T_{\rm C}$, using a lattice simulation with 2+1 dynamical quark
  flavours.  We used the {\sc hal qcd} time-dependent method,
  originally developed for inter-nucleon potentials.  Our lattices are
  anisotropic, with temporal lattice spacing less than the spatial one
  which enhances the information content of our correlators at each
  temperature. Local-extended charmonium correlators were calculated
  efficiently by contracting propagators in momentum rather than
  coordinate space. We find no significant variation in the central
  potential for temperatures in the confined phase. As the temperature
  increases into the deconfinement phase, the potential flattens,
  consistent with the expected weakening interaction.  We fit the
  potential to both the (a) Cornell and (b) Debye-screened potential
  forms, with the latter better reproducing the data.  The zero
  temperature string tension obtained from (a) agrees with results
  obtained elsewhere, and it decreases with temperature, but at a
  slower rate than from the static quark approximation.  The Debye
  mass from (b) is close to zero for small temperatures, but starts to
  increase rapidly around $T_{\rm C}$.  The spin-dependent potential
  is found to have a repulsive core and a distinct temperature
  dependence above $T_{\rm C}$ at distances $\sim 1$ fm.}

\begin{document} 
\maketitle
\flushbottom



\section{Introduction}
\label{sec:intro}

Shortly after the discovery of the $J/\psi$ particle, it was realised
that charmonium bound states could be very well described using the
Schr\"odinger equation with a phenomenological potential between the
quark and antiquark
\cite{Eichten:1974af,Eichten:1978tg,Eichten:1979ms}.  This was one of
the factors leading to QCD being accepted as the theory of the strong
interaction, with charmonium as the `hydrogen atom' of QCD.  The
validity of a potential model has since been formally established
using an effective theory --- potential nonrelativistic QCD (pNRQCD)
--- which is obtained by integrating out degrees of freedom with
momentum above the typical binding energy of the system
\cite{Brambilla:1999xf,Brambilla:2004jw}.  The resulting potential for
infinitely heavy (static) quarks can be shown to be equivalent to the
one extracted from the Wilson loop, which has been computed
non-perturbatively from first principles on the lattice. Recently, the
potential between quarks with {\em finite} mass has also been computed
from lattice QCD by `reverse engineering' the Schr\"odinger equation
\cite{Ikeda:2010nj,Ikeda:2011bs,Kawanai:2011xb,Kawanai:2011jt,Kawanai:2011fh,
  Kawanai:2013aca} using the {\sc hal qcd} method
\cite{Ishii:2006ec}.

At non-zero temperature, a potential model incorporating colour-Debye
screening has been used to predict the dissociation of charmonium
states as a signal for the formation of quark-gluon plasma (QGP)
\cite{Matsui:1986dk}.  Since that seminal paper, a substantial
experimental effort has been invested in the study of $J/\psi$
suppression at SPS, RHIC and the LHC \cite{Abreu:2000ni,
  Chatrchyan:2012np, Adcox:2004mh, Adams:2005dq}, and a number of
studies have been carried out using potential models for charmonium
systems at high temperature \cite{Mocsy:2007yj, Mocsy:2005qw,
  Mocsy:2004bv}.  However, unlike at zero temperature, there was no
rigorous proof of the validity of potential models for static quarks
and hence no agreement on what to use for the interquark potential.
Different groups have used, for example, the free energy
\cite{Kaczmarek:2005ui, Fodor:2007mi} or the internal energy
\cite{Kaczmarek:2005gi} of static quarks as computed on the lattice,
or a combination of these \cite{Wong:2004zr,Wong:2005be}.

Recently, a series of effective theories has been developed, depending
on hierarchies such as
\begin{equation}
M_q \gg T > g^2M_q > gT \gg g^4M_q\quad\text{or}\quad
M_q \gg g^2M_q > T \gg gT \gg g^4M_q,
\end{equation}
where $M_q$ is the heavy quark mass, $T$ is the temperature and $g$ is
the gauge coupling. A common feature of these theories is the
appearance of an imaginary part of the potential, resulting from
Landau damping \cite{Laine:2006ns,Laine:2007qy}.  In certain parameter
ranges this term can be more important for charmonium suppression than
the Debye screening encoded in the real part.

As in the zero temperature case, non-perturbative (lattice)
calculations of the Wilson loop have been used to extract the static
inter-quark potential at non-zero temperature \cite{Rothkopf:2011db,
  Burnier:2012az, Bazavov:2012bq, Bazavov:2012fk}.

In this work we further extend these non-perturbative lattice
calculations to the finite quark mass case. We will assume the
validity of the Schr\"odinger equation and potential description for
charm quarks, and derive the charmonium potential at non-zero
temperature directly from charmonium correlators following the {\sc
  hal qcd} method.  We restrict ourselves to the real part of the
potential and will be following the `time-dependent' method introduced
in \cite{Aoki:2012tk}. This method was used in \cite{Evans:2013yva}
but is distinct from the `fitting' method used in
\cite{Allton:2013wza}.  In the fitting method, local-extended
correlators are first fitted to exponentials at large Euclidean time,
$\tau$, to extract the Nambu-Bethe-Salpeter (NBS) ground state wave
function. The NBS wave function is then used, in conjunction with the
Schr\"odinger equation, to reverse-engineer the potential. The fitting
method is well understood from a theoretical point of view since it
relies on conventional fitting techniques. However, at non-zero
temperature, where the temporal range of the correlators is limited,
it suffers from familiar limitations: higher excited states still
contribute to the correlator at the largest available $\tau$, making
fits unreliable.

In \cite{Iida:2011za} the first calculation of the charmonium
potential at finite mass and non-zero temperature was performed, but
this used the quenched approximation. We performed thermal studies of
the charmonium potential using our two-flavour, 1st generation {\sc
  fastsum} ensembles in our previous study
\cite{Evans:2013yva,Allton:2013wza}.  In the work presented here, we
extend this by using our 2nd generation ensembles which have 2+1
flavours with finer, larger lattices and light quarks closer to their
physical masses.  We find that the potential does not vary
significantly for temperatures below the crossover temperature, $T_C$, but
that it clearly flattens above $T_C$.  Using the Cornell form of the
potential, we determine the temperature dependent string tension which
decreases as $T$ increases as expected. We also fit our data to a
Debye-screened potential form to determine the Debye mass. The
spin-dependent potential is also determined and we find thermal
effects for $T\gtap T_C$. An early version of this work appears in
\cite{Evans:2013zca}.

In section \ref{sec:hal_method}, the {\sc hal qcd} time-dependent
method is reviewed, and we discuss the effect of the backward moving
states. Our simulation details are outlined in section \ref{sec:simulation}
and section \ref{sec:results} presents our results.



\section{The Method}
\label{sec:hal_method}


\begin{figure}[t]
\centering
\includegraphics[scale=1,trim=80 560 0 120,clip]{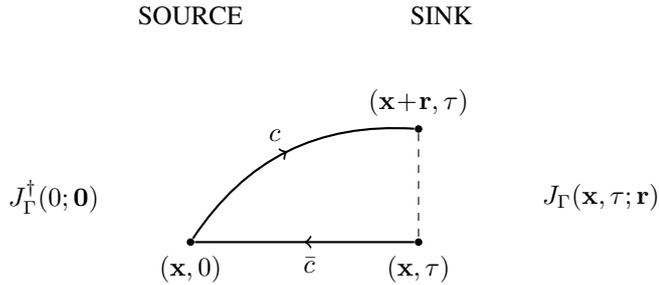}
\caption{Quark propagator diagram of a local-extended correlator. The
  quark and anti-quark are created at the source with no separation
  and annihilated at the sink with a separation $\mathbf{r}$.}
\label{fig:locext}
\end{figure}


The {\sc hal qcd} time-dependent method of extracting the potential
from the lattice \cite{Aoki:2012tk} takes local-extended correlators
as input, see Figure \ref{fig:locext}. These are constructed from
creation and annihilation operators that have the form
\begin{equation}
J_\Gamma(x;\mathbf{r}) = \bar{q}(x)\,\Gamma\, U(x,x\!+\!\mathbf{r})\,q(x\!+\!\mathbf{r}),
\label{eq:J}
\end{equation}
where $\mathbf{r}$ is the separation between the quark and anti-quark
fields $q$ and $\bar{q}$, $x$ is the space-time point
$(\mathbf{x},\tau)$ and $\Gamma$ is a monomial of gamma matrices used
to generate pseudoscalar ($\eta_c$) channels or vector ($J/\psi$).
$U(x,x\!+\!\mathbf{r})$ is the gauge connection between $x$ and
$x+\mathbf{r}$ required for gauge invariance.

The local-extended charmonium correlator is
\begin{equation} 
C_\Gamma(\mathbf{r},\tau) =
\sum_{\mathbf{x}} \langle J_\Gamma(\mathbf{x},\tau;\mathbf{r}) \; J_\Gamma^\dagger(0;\mathbf{0}) \rangle,
\label{eq:le_cf}
\end{equation}
where the sum over the spatial coordinate at the sink, $\mathbf{x}$,
projects the momentum of the state to zero.

The local-extended correlator can also be expressed as a sum over the
eigenstates of the Hamiltonian with eigenvalues $E_j$,
\begin{equation}
C_\Gamma(\mathbf{r},\tau) = \sum_j \frac{\psi_j^\ast(\mathbf{0}) \psi_j(\mathbf{r})}{2E_j}
\;\left( e^{-E_j \tau} + e^{-E_j (N_\tau - \tau)} \right),
\label{eq:cfn}
\end{equation}
where the sum is over the states $j$ with the same Lorentz
transformation properties as the operator $J_\Gamma$, $N_\tau$ is the
number of lattice points in the temporal direction and
$\psi_j(\mathbf{r})$ is the charmonium wave function.  We now consider
only the forward-moving contribution to the correlator (the effect of
ignoring the backward-moving contribution is discussed later):
\begin{equation}
C_\Gamma(\mathbf{r},\tau) = \sum_j \frac{\psi_j^\ast(\mathbf{0})
  \psi_j(\mathbf{r})}{2E_j}e^{-E_j \tau}\equiv\sum_j
\Psi_j(\mathbf{r})e^{-E_j \tau},
\label{eq:forward_cfn}
\end{equation}
where we have defined the unnormalised wavefunction,
$\Psi_j(\mathbf{r})=\psi_j^\ast(\mathbf{0})\psi_j(\mathbf{r})/2E_j$.
We treat the charm quark non-relativistically due to its large mass,
and assume $\Psi_j(\mathbf{r})$ obeys the Schr\"odinger equation,
\begin{equation}
\left(- \frac{\nabla^2_r}{2\mu} + V_\Gamma(r)
  \right) \Psi_j(r)
= E_j \Psi_j(r),
\label{eq:hal_schro}
\end{equation}
where $V_\Gamma(r)$ is the desired potential for the channel $\Gamma$,
$\mu = m_c/2$ is the reduced mass, and we only consider S-wave states.
Taking the time derivative of \eqref{eq:forward_cfn} and using
\eqref{eq:hal_schro}, we obtain,
\begin{eqnarray} \nonumber
\frac{\partial C_\Gamma(\mathbf{r},\tau)}{\partial \tau}
= -\sum_j E_j \Psi_j(\mathbf{r})e^{-E_j \tau}
&=& \sum_j \left( \frac{\nabla^2_r}{2\mu} - V_\Gamma(r) \right) \Psi_j(r) e^{-E_j \tau} \\
&=& \left( \frac{\nabla^2_r}{2\mu} 
          - V_\Gamma(r) \right) C_\Gamma(\mathbf{r},\tau)
\label{eq:V}
\end{eqnarray}
which can be trivially rearranged to yield the potential,
\begin{equation}
V_\Gamma(r) = \frac{1}{C_\Gamma(\mathbf{r},\tau)}
\left( \frac{\nabla^2_r}{2\mu} - \frac{\partial}{\partial \tau} \right) C_\Gamma(\mathbf{r},\tau).
\label{eq:HALQCD_pot}
\end{equation}
We highlight the fact that $V_\Gamma(r)$ from \eqref{eq:HALQCD_pot}
has an implicit $\tau$ dependence which must be averaged over, see
section \ref{sec:results}.

On the lattice, the Laplacian in \eqref{eq:HALQCD_pot} is approximated as follows,
\begin{align}
\nabla_{r}^2f(r)
\equiv\left[ \frac{\partial^2}{\partial r^2} + \frac{2}{r}\frac{\partial}{\partial r}\right] f(r)
\longrightarrow
\left[\frac{\delta_{r',r+a_s} - 2\delta_{r',r} + \delta_{r',r-a_s}}{a_s^2}
 + \frac{\delta_{r',r+a_s} - \delta_{r',r-a_s}}{ra_s}\right]f(r'),
\label{eq:laplacian}
\end{align} 
where we have relied on the approximate rotational symmetry of the
lattice.

The time derivative in \eqref{eq:HALQCD_pot} can be approximated by
the naive finite temporal difference,
\begin{align}
\frac{\partial}{\partial\tau} f(\tau) \longrightarrow
\left[ \frac{f(\tau + a_\tau) - f(\tau - a_\tau)}{2a_\tau}\right],
\label{eq:time_der}
\end{align}
but, as we will see in section \ref{sec:central}, this
approximation is particularly poor near the temporal centre of the
lattice because of contamination by the backward mover which we
neglected in the above derivation going from \eqref{eq:cfn} to
\eqref{eq:forward_cfn}.  For large temperatures, corresponding to
lattices with a small temporal extent, this is especially problematic
because the uncontaminated region may become vanishingly small.

Fortunately the expression \cite{Durr:2012te}
\begin{align}
\tilde{E_0}(\tau) = \frac{1}{2}\log
\left(
\frac
{C_\Gamma(\tau-a_\tau)+\sqrt{C_\Gamma(\tau-a_\tau)^2-C_\Gamma(N_\tau a_\tau/2)^2}}
{C_\Gamma(\tau+a_\tau)+\sqrt{C_\Gamma(\tau+a_\tau)^2-C_\Gamma(N_\tau a_\tau/2)^2}}
\right),
\label{eq:backward}
\end{align}
recovers the exact ground state energy, $E_0$, even in the case where
there is a backward mover as in \eqref{eq:cfn} (in the absence of
excited states).

In the next section, we test the improved temporal difference based on
\eqref{eq:backward},
\begin{align}
\frac{\partial}{\partial\tau} C_\Gamma(\tau) \longrightarrow
-\tilde{E_0} C_\Gamma(\tau)
\label{eq:time_der2}
\end{align}
to define the potential, \eqref{eq:HALQCD_pot}. We find it has
significantly reduced contamination from the backward mover compared
to the naive expression \eqref{eq:time_der}.

Once the potential for both the pseudoscalar (PS) and vector (V)
channels have been determined from the method described above, the
central and spin-dependent potentials, $V_{\rm C}$ and $V_{\rm S}$,
can be derived as follows. The leading order terms in the velocity
expansion of the interquark potential for S-wave states can be
expressed as \cite{Godfrey:1985xj,Barnes:2005pb},
\begin{align}
V_\Gamma(\mathbf{r})=V_{\rm C}(\mathbf{r})+V_{\rm S}(\mathbf{r})\,s_1\cdot s_2,
\end{align}
where $s_{1,2}$ are the quark spins. Using $s_1\cdot s_2 =-3/4,\;1/4$
for the PS and V channels respectively, $V_{\rm C}(\mathbf{r})$ and
$V_{\rm S}(\mathbf{r})$ can be obtained from the $V_{\rm PS, V}$
potentials using the following expressions,
\begin{equation}
V_{\rm C}(\mathbf{r})=\frac{1}{4}V_{\rm PS}+\frac{3}{4}V_{\rm V}, 
\quad\quad
V_{\rm S}(\mathbf{r})=V_{\rm V}-V_{\rm PS}.
\label{eq:V_CS}
\end{equation}



\section{Simulation Details}
\label{sec:simulation}


\begin{table}
\label{tab:params}
\begin{center}
\begin{tabular}{ccccr}
\hline
 \multicolumn{1}{c}{$N_s$} & \multicolumn{1}{c}{$N_\tau$} & \multicolumn{1}{c}{$T$(MeV)} &
\multicolumn{1}{c}{$T/T_c$} & \multicolumn{1}{c}{$N_{\rm cfg}$} \\
\hline 
24 & 128 & 44 & 0.24 & 250 \\   
24 & 40 & 141 & 0.76 & 500 \\      
24 & 36 & 156 & 0.84 & 500 \\ 
24 & 32 & 176 & 0.95 & 1000 \\ 
24 & 28 & 201 & 1.09 & 1000 \\ 
24 & 24 & 235 & 1.27 & 1000 \\ 
24 & 20 & 281 & 1.52  & 1000 \\
24 & 16 & 352 & 1.90 & 1000 \\
\hline
32 & 32 & 176 & 0.95  & 500 \\
32 & 24 & 235 & 1.27  & 500 \\ 
\hline
\end{tabular}
\caption{The spatial and temporal extents, $N_s$ and $N_\tau$, the
  temperature in units of MeV and T$_{\rm C}$, and the number of
  configurations, $N_{\rm cfg}$, of the ensembles used. Note that for
  two temperatures, we have two different spatial volumes to study
  finite volume effects.}
\end{center}
\end{table}


We use our {\sc fastsum} Collaboration's 2nd generation configurations
for this analysis \cite{Allton:2014uia}.  A Symanzik gauge action and
a stout-smeared clover fermion action were used with parameters set by
the Hadron Spectrum Collaboration ({\sc hsc}) in
\cite{Edwards:2008ja,Lin:2008pr}.  There are 2+1 flavours of dynamical
quarks with the two (degenerate) light flavours corresponding to
$M_\pi = 392(4)$~MeV and the third dynamical quark set to the strange
quark mass.  Anisotropic lattices were employed with an anisotropy of
$\xi \equiv a_s / a_\tau \sim 3.5$, with $a_s\simeq0.123$~fm and
$a_\tau^{-1} \simeq 5.63${\ }GeV.  The lattice spacings are fixed and
we vary the temperature, $T=(a_\tau N_\tau)^{-1}$, by adjusting the
number of temporal points, $N_\tau$.  The ensemble parameters are
listed in Table \ref{tab:params} showing that we study temperatures in
both the confined and deconfined phases.  The ``zero'' temperature
(i.e. $N_\tau=128$) ensemble was kindly provided by the {\sc hsc}
Collaboration.  Our main spatial volume is $(3\text{fm})^3$,
corresponding to $N_s=24$, and we have two temperatures with a
$(4\text{fm})^3$ volume ($N_s=32$) to enable us to check finite volume
effects.  The deconfinement crossover temperature, $T_{\rm C}$, was
determined from the inflection point of the Polyakov loop
\cite{Allton:2014uia,Aarts:2014nba}.

The (non-dynamical) charm quark was calculated with the
same (relativistic) action used for the three light dynamical quarks
with its mass set by tuning the PS state to the experimental $\eta_c$
mass while simultaneously maintaining the anisotropy
\cite{Liu:2012ze}. As in \cite{Ikeda:2011bs}, the quark
mass is defined as $M_{V}/2$ where $M_V$ is the mass of the charmonium
vector channel ground state. Hence the reduced mass, $\mu=M_V/4$,
see \eqref{eq:hal_schro}.

We chose to gauge fix our configurations to the Coulomb gauge, and
then replace the gauge connection, $U(x,x+\mathbf{r})$, in
\eqref{eq:J} by unity.  We used the highly optimised
Fourier-accelerated gauge fixing procedure of \cite{Hudspith:2014oja}.
The coordinate space quark propagators were calculated using the
Chroma software suite \cite{Edwards:2004sx} and then tied together in
momentum space using a bespoke program to obtain correlators more
efficiently, see the Appendix for details.



\section{Results}
\label{sec:results}


\subsection{Central Potential}
\label{sec:central}

\begin{figure}
\centering
\includegraphics[scale=0.7, trim = 0 420 0 0,clip=true]{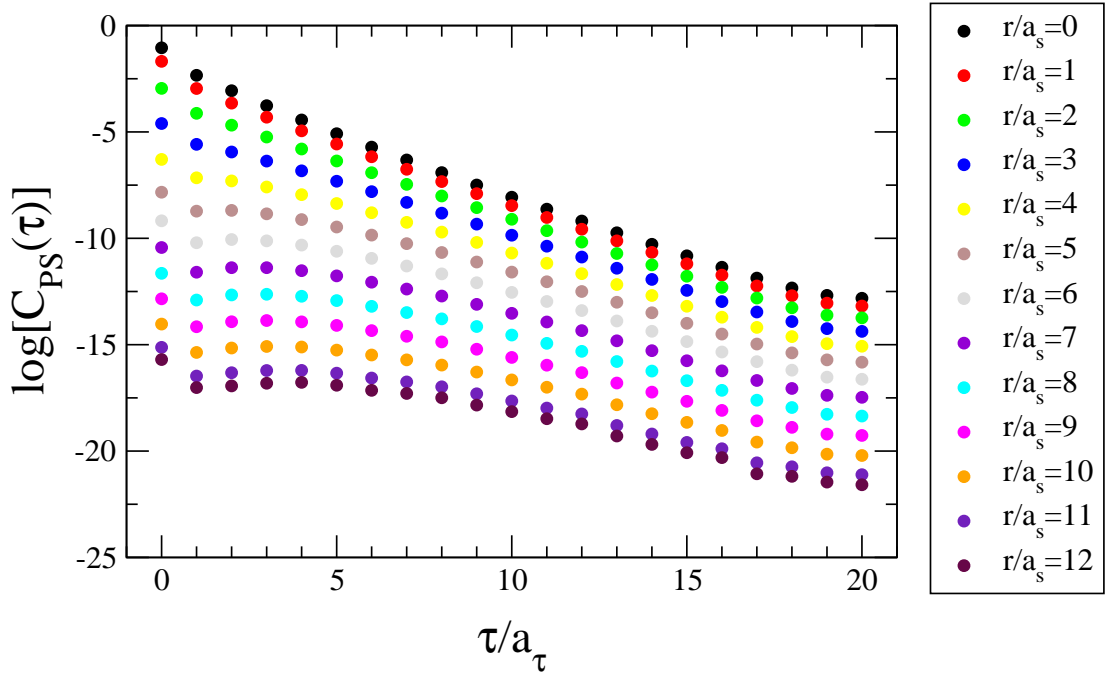}
\caption{Local-extended charmonium correlators for the PS channel for
  all possible on-axis separations of the $0.76T_c$ ($N_\tau=40$)
  ensemble. Error bars are smaller than the symbols.}
\label{fig:axis_Ct}
\end{figure}

Local-extended correlators, \eqref{eq:le_cf}, corresponding to on-axis
quark separations were generated from the ensembles outlined in
section \ref{sec:simulation}. Correlators corresponding to quark
separations of the same magnitude, $|\mathbf{r}|$, were averaged
giving 13 unique separations for $N_s=24$. The set of PS correlators
for the 0.76T$_{\rm C}$ ensemble is shown in Figure \ref{fig:axis_Ct}
for all available $\mathbf{r}$. As one would expect, the signal
decreases as the quark separation, $\mathbf{r}$, increases.  Assuming
ground state dominance of the correlator, this follows from the
monotonic property of the S-wave wavefunction.


\begin{figure}[]
\centering
\includegraphics[scale=0.7,trim = 0 425 0 5, clip=true]{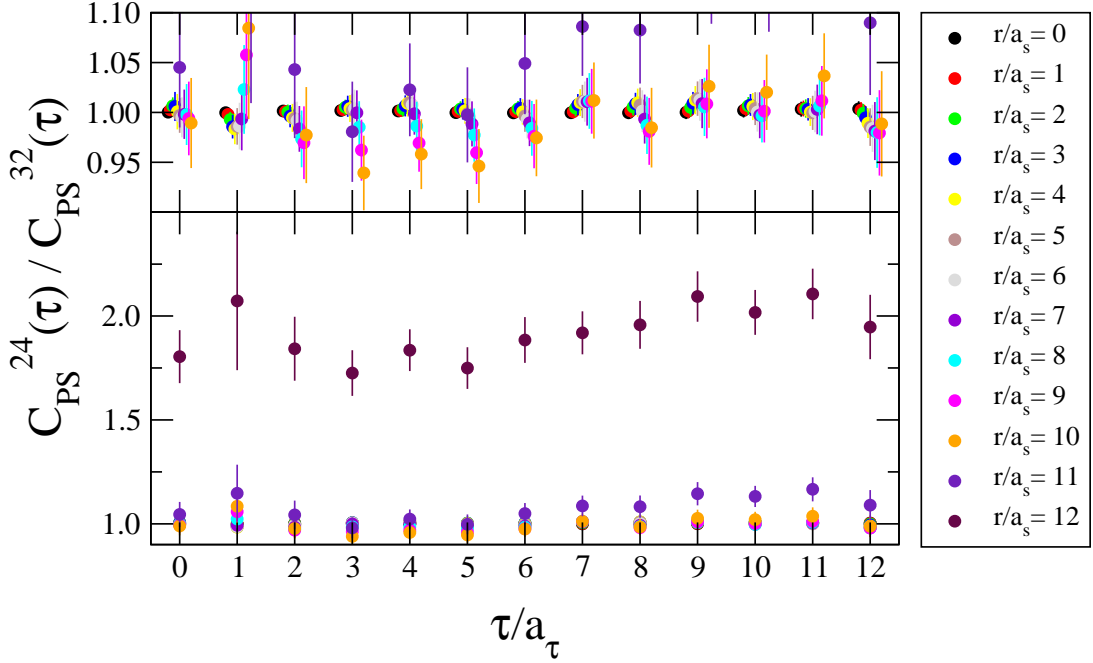}
\caption{The ratio of the $N_s=24$ to $N_s=32$
  local-extended correlation functions for $1.27T_c$
  (i.e. $N_\tau\!=\!24$).  In the lower pane, all separations
  $0\le|\mathbf{r}/a_\tau|\le 12$ are shown.  The upper pane shows a
  closeup for $0\le|\mathbf{r}/a_\tau|\le 10$ with the points shifted
  horizontally for clarity.}
\label{fig:PSaxisVol}
\end{figure}

The $N_s=32$ ensembles, listed in Table \ref{tab:params}, provide a
means to investigate the volume dependence of the correlators, and
hence also that of the potentials.  Figure \ref{fig:PSaxisVol} shows
the ratio of $N_s=24$ to $N_s=32$ local-extended correlators for the
$1.27T_c$ (i.e. the $N_\tau=24$) case.  From Figure \ref{fig:PSaxisVol},
the correlator has no volume dependence for $0 \le r/a_s \le 10$, the
$r/a_s=11$ case shows some effects and the $r/a_s=12$ is clearly
highly sensitive to finite volume effects.  Consequently, (due to the
nearest-neighbour representation of the Laplacian in
\eqref{eq:laplacian}) we report the potential for $r/a_s\le 9$,
i.e. $r \le 1.1$fm, in the following where it is free from finite
volume effects.

We note that the lattice version of the Laplacian in
\eqref{eq:laplacian} has greatest discretisation error near the
origin. Hence we do not include the $r/a_s=1$ point in our fits to the
potential in section \ref{sec:string}.


\begin{figure}
\centering
\includegraphics[scale=0.7, trim = 0 425 0 0, clip=true]{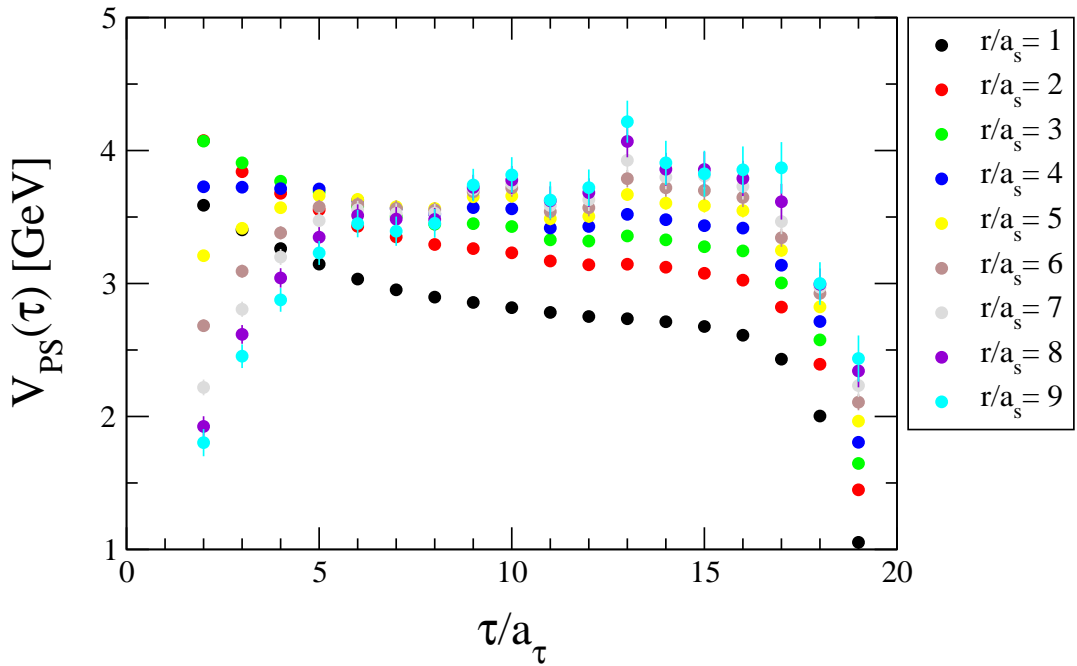}
\caption{The potential for the $0.76 T_{\textrm{C}}$
  pseudoscalar channel plotted as a function of $\tau$ for each quark
  separation $|\mathbf{r}|$, using \eqref{eq:HALQCD_pot} with the
  naive time derivative \eqref{eq:time_der}.}
\label{fig:40_PS_timeslice_pot_raw}
\end{figure}

\begin{figure}
\centering
\includegraphics[scale=0.7, trim = 0 425 0 0, clip=true]{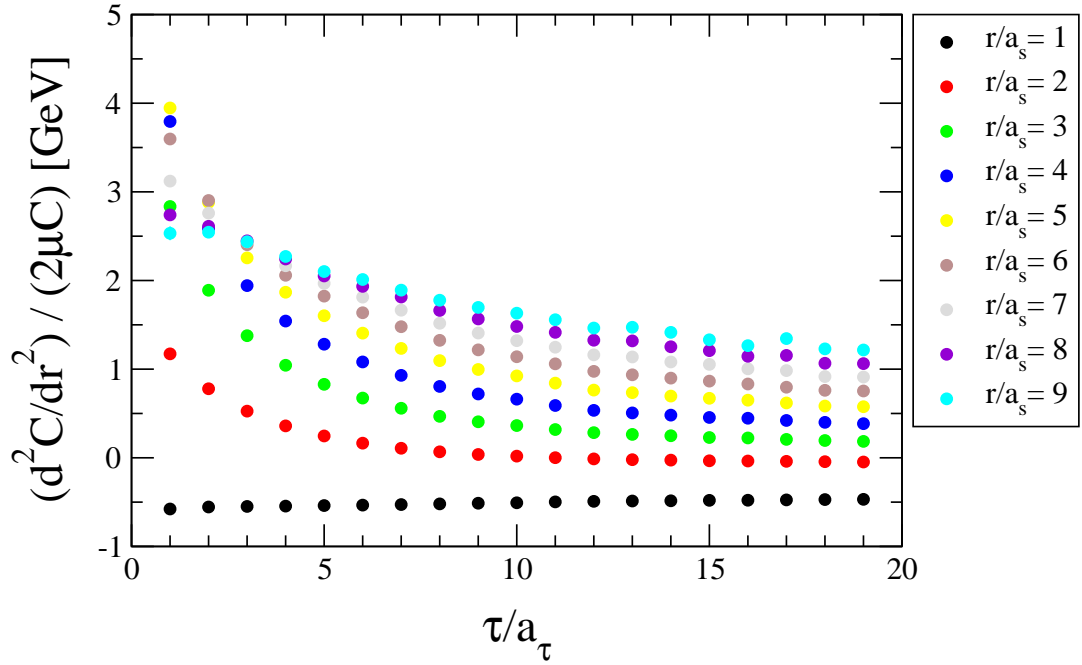}
\caption{The spatial contribution, $(1/C_\Gamma)(\nabla_r^2
  C_\Gamma/2\mu)$, to the potential (see
  \eqref{eq:HALQCD_pot}) for the $0.76T_{\textrm{C}}$ pseudoscalar
  channel, plotted as a function of $\tau$ for each quark separation
  $|\mathbf{r}|$.}
\label{fig:40_PS_timeslice_1st_Term}
\end{figure}

\begin{figure}
\centering
\includegraphics[scale=0.7, trim = 0 425 0 0, clip=true]{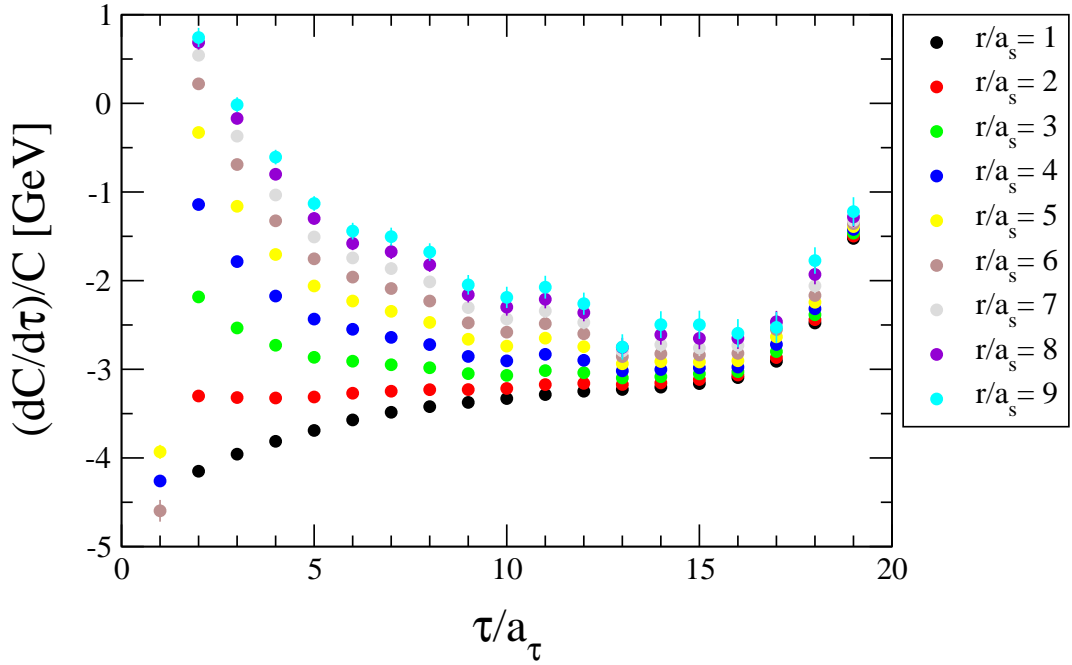}
\caption{The temporal contribution, $(1/C_\Gamma)(\partial
  C_\Gamma/\partial\tau)$, to the potential (see
  \eqref{eq:HALQCD_pot}), using the naive form, \eqref{eq:time_der},
  for the $0.76$T$_{\textrm{C}}$ pseudoscalar channel, plotted as a
  function of $\tau$ for each quark separation $|\mathbf{r}|$.}
\label{fig:40_PS_timeslice_2nd_Term}
\end{figure}

\begin{figure}
\centering
\includegraphics[scale=0.7, trim = 0 425 0 0, clip=true]{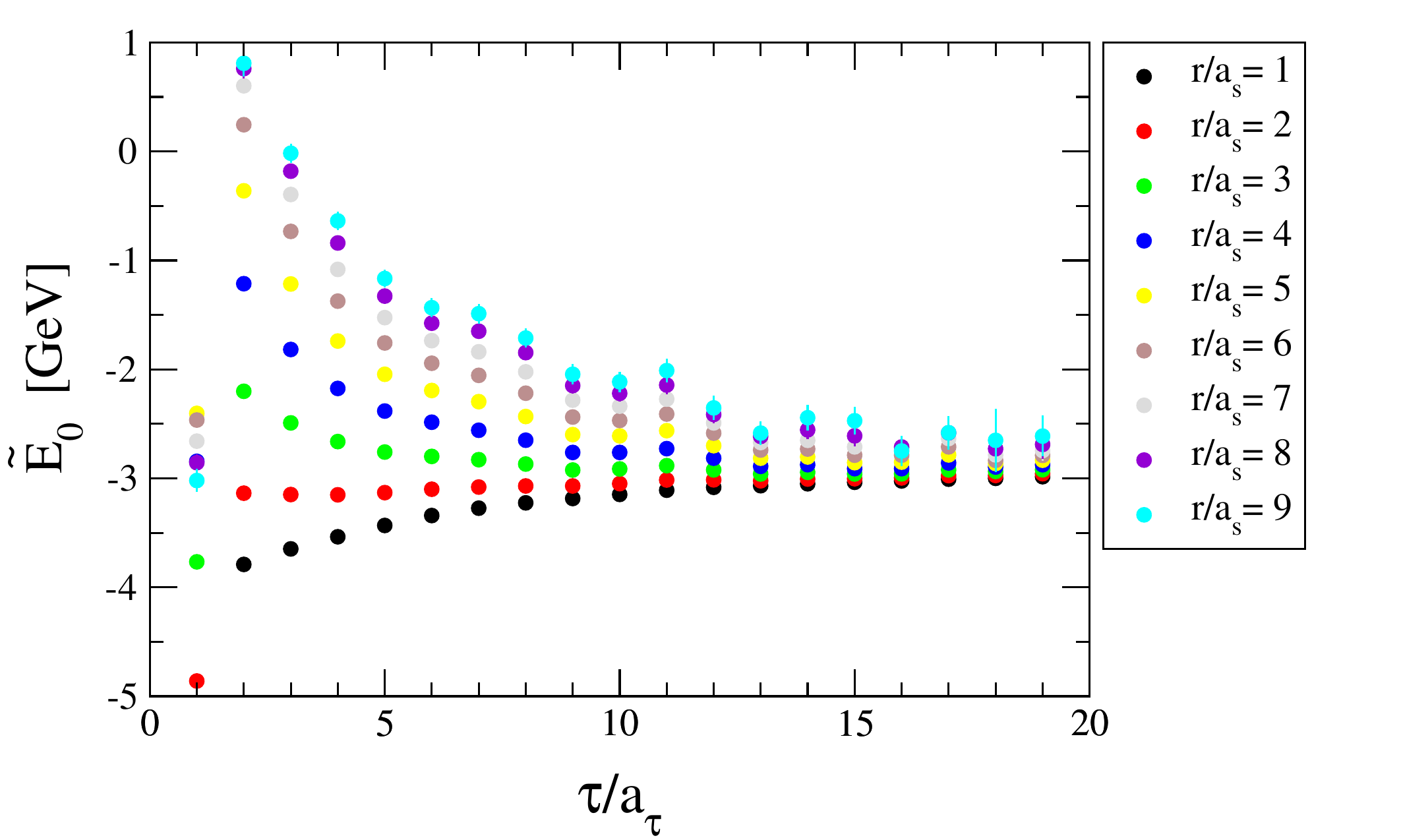}
\caption{The temporal contribution, $(1/C_\Gamma)(\partial
  C_\Gamma/\partial\tau)$, to the potential, using the improved
  temporal form, $\tilde{E_0}$, see \eqref{eq:backward}, for the
  $0.76$T$_{\textrm{C}}$ pseudoscalar channel for each quark
  separation $|\mathbf{r}|$.}
\label{fig:40_PS_timeslice_aM_Term}
\end{figure}

In Figure \ref{fig:40_PS_timeslice_pot_raw} the PS potential is
plotted for the 0.76T$_{\rm C}$ correlators using the naive time
derivative \eqref{eq:time_der} in \eqref{eq:HALQCD_pot}.  We note that
in the {\sc hal qcd} method, the potential has a $\tau$
pseudo-dependence which will be averaged over. Figures
\ref{fig:40_PS_timeslice_1st_Term} and
\ref{fig:40_PS_timeslice_2nd_Term} show the spatial and temporal
contributions of \eqref{eq:HALQCD_pot} respectively.  As can been
seen, at large $\tau$ the spatial derivative contribution is stable
but the temporal derivative contribution increases near the mid-point
of the lattice, $\tau\sim N_\tau/2$, which produces a corresponding
decrease in the potential at these $\tau$ values in Figure
\ref{fig:40_PS_timeslice_pot_raw}.  In Figure
\ref{fig:40_PS_timeslice_aM_Term} we use the improved temporal term
\eqref{eq:backward} which clearly resolves the issue for the
pseudoscalar case, implying that it was caused by the backward mover
as discussed in section \ref{sec:hal_method}. We have checked that the
same is true in the vector channel.  We also note that as
$\tau\rightarrow \infty$ there is the expected convergence towards the
(negative) value of the PS mass $\eta_c=2.9804(1)$GeV
\cite{Beringer:1900zz}.  We note that in the region where
\eqref{eq:HALQCD_pot} is valid, i.e. in the absence of a backward
mover, $\tilde{E_0}$ is equivalent to $\partial C_\Gamma/\partial\tau$
only in the limit of negligible excited state contribution. However,
noting the very similar small-$\tau$ behaviour in Figures
\ref{fig:40_PS_timeslice_2nd_Term} and
\ref{fig:40_PS_timeslice_aM_Term} (where the excited states'
contributions are largest), it is clear that the $\tilde{E_0}$
definition works in practice even in the presence of excited states.
For these reasons, we will always use the improved temporal form
\eqref{eq:backward} in the following.


\begin{figure}
\centering
\includegraphics[scale=0.71, trim = 0 0 0 0, clip=true]{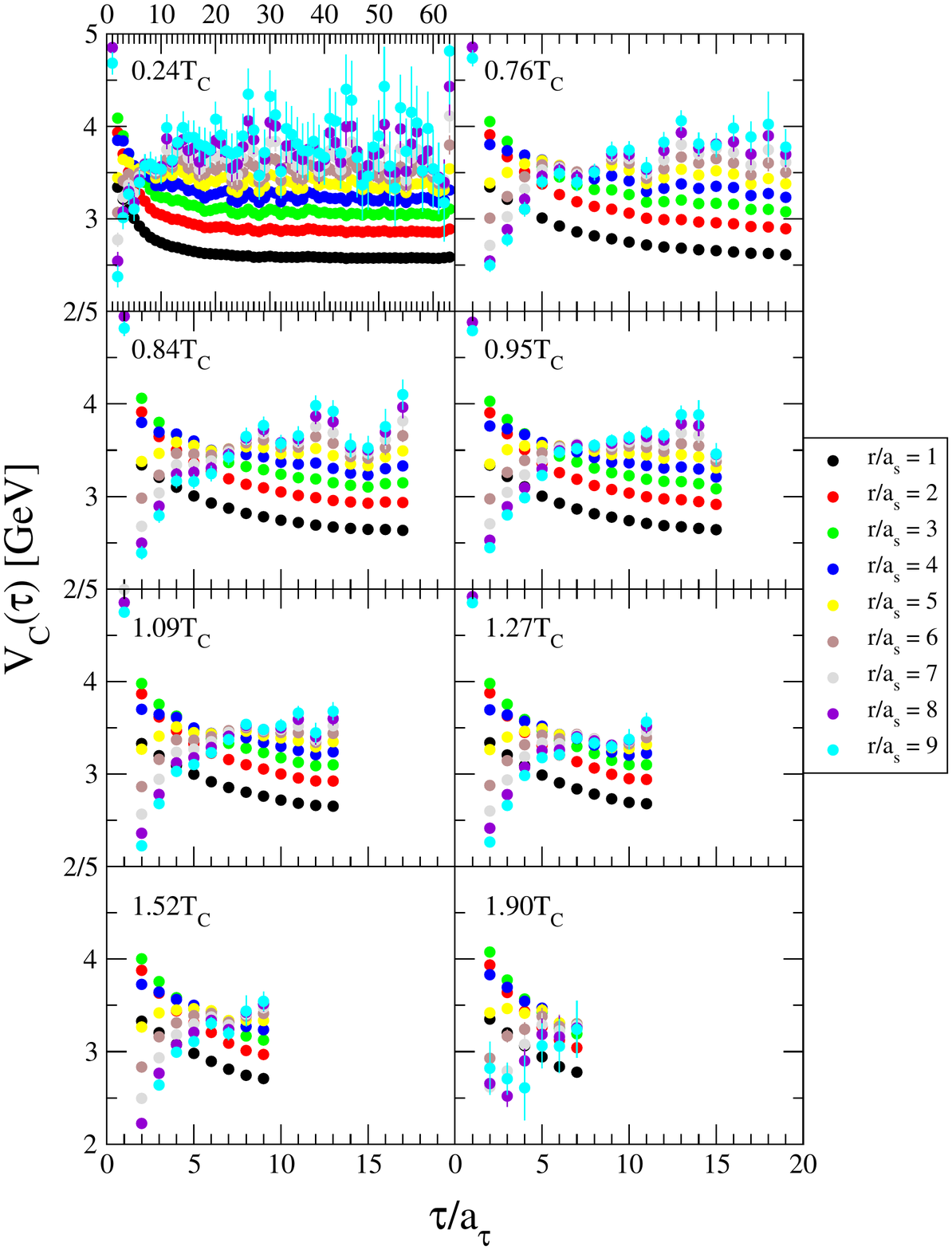}
\caption{The central potential for all temperatures studied as a
  function of $\tau$ for each quark separation, $|\mathbf{r}|$.}
\label{fig:Vc_allT}
\end{figure}

The central potential is determined by combining the pseudoscalar and
vector potentials according to \eqref{eq:V_CS} and is shown in Figure
\ref{fig:Vc_allT} for all temperatures.  From these figures, it is
clear that the 1.52T$_{\rm C}$ and 1.90T$_{\rm C}$ data do not
stabilise. Hence, we include only the data up to temperatures of
1.27T$_{\rm C}$ in the following.


\begin{table}[h]
\centering
\begin{tabular}{cccc}
\hline
$T/T_{\rm C}$ & $N_\tau$ & Best range & Lower range \\
\hline 
0.24   & 128 & $30-63$ & $15-19$ \\
0.76   &  40 & $15-19$ & $12-17$ \\
0.84   &  36 & $12-17$ & $11-15$  \\
0.95   &  32 & $11-15$ & $11-13$  \\
1.09   &  28 & $11-13$ & $9-11$  \\
1.27   &  24 & $9-11$  & N/A  \\
\hline
\end{tabular}
\caption{Fitting ranges used to
  obtain the $\tau-$independent potentials. The ``best'' range
  covers the plateau region for each temperature and gives our best
  fit. The ``lower'' range is the same as the best range for the
  next higher temperature, allowing a direct comparison between
  temperatures.
}
\label{tab:nf2plus1cratau}
\end{table}

\begin{figure}
\centering
\includegraphics[scale=0.7, trim = 0 360 0 0, clip=true]{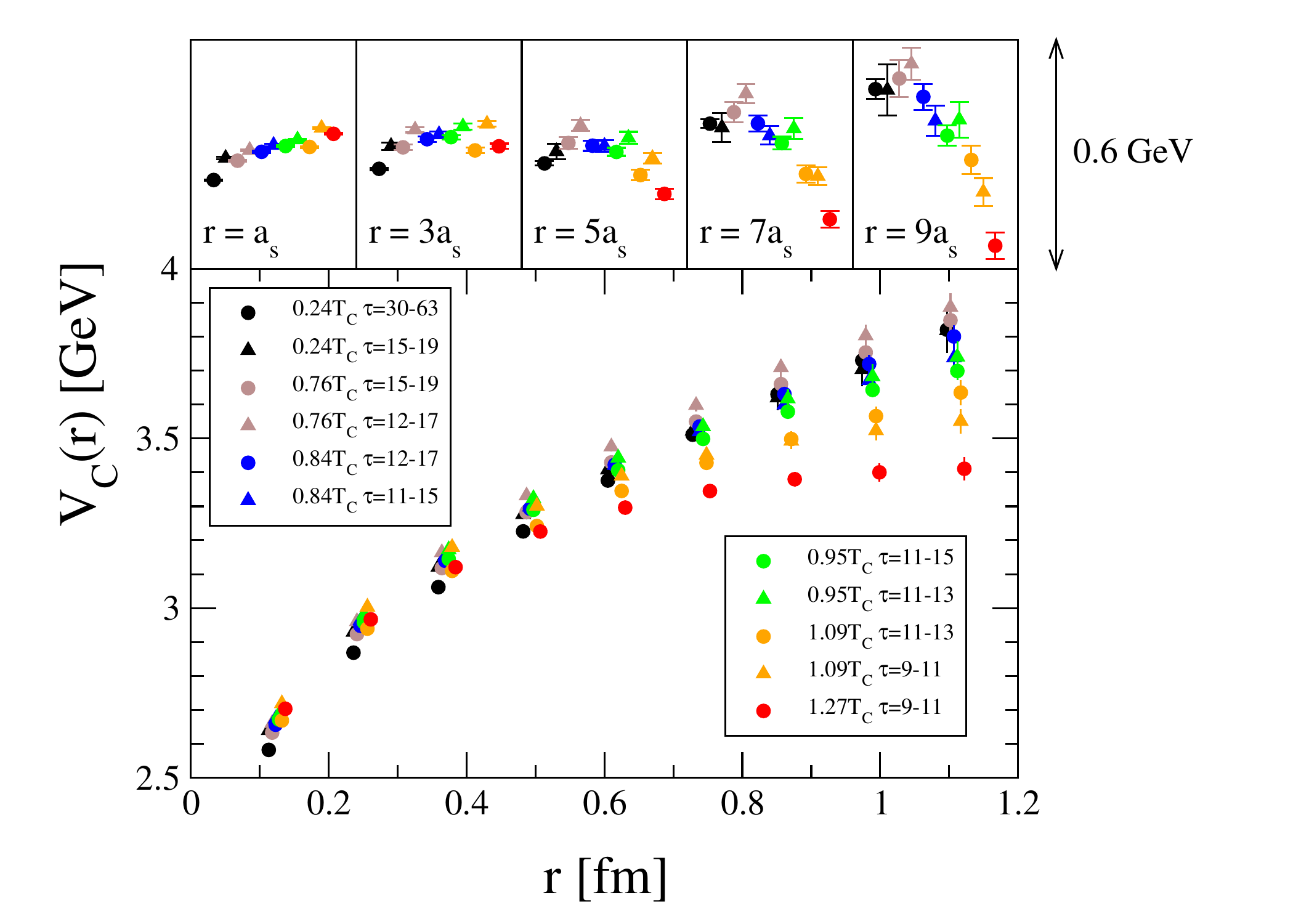}
\caption{The central potential for
  various temperatures as a function of separation. Two different
  $\tau$ ranges were used to estimate the systematic error for all but
  the largest temperature, see Table \ref{tab:nf2plus1cratau}. The
  circles represent the ``best'' fit range in $\tau$ for each
  temperature, and the triangles the ``lower'' fit range (which is the
  next highest temperature's best range). This allows a direct
  comparison at different temperatures to be made. The five upper
  graphs are closeups of the separations $r/a_s = 1,3, \ldots 9$ with
  a common vertical range of 0.6 GeV. All points are shifted slightly
  horizontally for clarity.}
\label{fig:Nf2plus1_Vc_aM_all_Ntau}
\end{figure}

We remove the $\tau$ pseudo-dependence from the potentials by
performing a correlated fit to a constant using the $\tau$ ranges shown in Table
\ref{tab:nf2plus1cratau}. Two time ranges are chosen to elucidate the
systematic error from the choice of fit range. The ``best'' range
is chosen to be the best available for each temperature, and the
``lower'' range is chosen to match the best range of the next
higher temperature. In this way, direct comparisons can be made
between neighbouring temperatures.  Using these time ranges, we obtain
the temperature-dependent potential shown in Figure
\ref{fig:Nf2plus1_Vc_aM_all_Ntau}. The circles correspond to the
best range and triangles the lower range. To aid this comparison
between temperatures, we include five upper insert plots in Figure
\ref{fig:Nf2plus1_Vc_aM_all_Ntau} which show a closeup of the data at
every second separation, i.e. $r = a_s, 3a_s, \ldots 9a_s$. The
vertical range of all the insert plots is identical (0.6 GeV). We
emphasize from the discussion above, a direct comparison between
neighbouring temperatures can be made by comparing the triangles at one
temperature with the circles at the next higher temperature. We
therefore conclude that the $T \ltap T_{\rm C}$ potentials have no
significant temperature effects that our data can discern for any
separation, but that there is a significant flattening of the
potential at moderate to large separations, $r\gtap 0.3$ fm, for $T
\gtap T_{\rm C}$.

In Table \ref{tab:pot}, the central potential values are
listed with the systematic (from the $\tau$ fitting procedure
described above)
and statistical errors combined additively into a single
error bar.


\begin{table}[h]
\begin{center}
\begin{tabular}{c|cccccc}
\hline
&&&&&\\
$r$ [fm] & \multicolumn{5}{c}{$V_C$ [GeV]} \\
& $0.24\;T_C$ & $0.76\;T_C$ & $0.84\;T_C$ & $0.95\;T_C$ & $1.09\;T_C$ & $1.27\;T_C$ \\
&&&&&\\
\hline
&&&&&\\
 0.123 & 2.58(6) & 2.63(3) & 2.66(2) & 2.67(2) & 2.67(5) & 2.703(2) \\
&&&&&\\
 0.246 & 2.87(6) & 2.92(4) & 2.95(2) & 2.96(2) & 2.94(6) & 2.967(5) \\
&&&&&\\
 0.369 & 3.06(6) & 3.12(5) & 3.14(1) & 3.15(3) & 3.11(7) & 3.121(7) \\
&&&&&\\
 0.492 & 3.23(5) & 3.28(5) & 3.29(1) & 3.29(3) & 3.24(6) & 3.23(1) \\
&&&&&\\
 0.615 & 3.38(3) & 3.43(5) & 3.42(1) & 3.41(4) & 3.35(4) & 3.30(1) \\
&&&&&\\
 0.738 & 3.51(1) & 3.55(5) & 3.54(2) & 3.50(4) & 3.43(2) & 3.35(2) \\
&&&&&\\
 0.861 & 3.63(1) & 3.66(5) & 3.63(3) & 3.58(4) & 3.50(2) & 3.38(2) \\
&&&&&\\
 0.984 & 3.73(3) & 3.75(5) & 3.72(5) & 3.64(4) & 3.57(4) & 3.40(3) \\
&&&&&\\
 1.107 & 3.82(3) & 3.85(5) & 3.80(6) & 3.70(4) & 3.64(8) & 3.41(3) \\
&&&&&\\
\hline
\end{tabular}
\caption{The central potential data as shown in Figure
  \ref{fig:Vc_with_Cornell}.  The statistical uncertainty has been
  combined additively with the systematic error from the $\tau-$range
  choice as described in the text, but note that the errors for the
  $1.27T_{\rm C}$ data are statistical only.
\label{tab:pot}
}
\end{center}
\end{table}




\subsection{The Cornell Potential, String Tension and Debye Screening}
\label{sec:string}

\begin{figure}
\centering
\includegraphics[scale=0.7, trim = 0 420 0 0, clip=true]{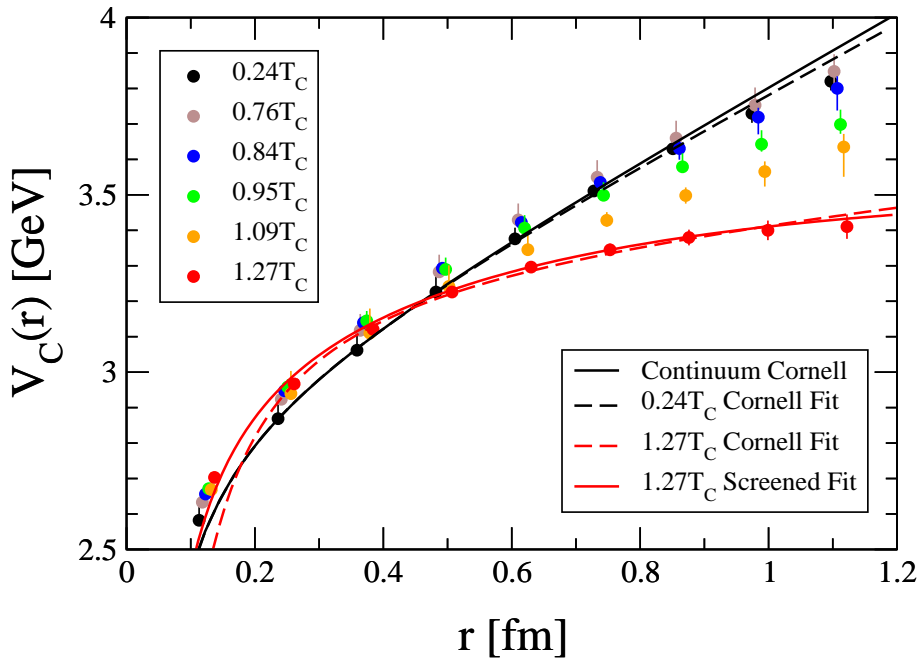}
\caption{The central potential with combined statistical and
  systematic errors for various temperatures (noting that the
  1.27$T_{\rm C}$ data has statistical errors only). Data points are
  shifted slightly horizontally for clarity. Examples of the fits to
  our data are shown using the Cornell \eqref{eq:cornell} and
  Debye-screened \eqref{eq:screened} potentials.  These used the
  restricted range $2a_s \le r \le 9a_s$ as described in the
  text. The Cornell form from \eqref{eq:cornell} with the
    continuum parameters $\alpha_{\rm c}=\pi/12$
    \cite{Luscher:1980ac} and $\sqrt{\sigma}=0.445$ GeV
    \cite{Satz:2005hx} is shown which agrees extremely well with
  the Cornell fit to our $0.24T_{\rm c}$ data.}
\label{fig:Vc_with_Cornell}
\end{figure}

In Figure \ref{fig:Vc_with_Cornell} the central potential is plotted
with the combined statistical and systematic errors as listed in Table
\ref{tab:pot}. For comparison, the Cornell potential
\cite{Eichten:1974af,Eichten:1978tg,Eichten:1979ms},
\begin{equation}
V(r,T) = -\frac{\alpha_{\rm c}(T)}{r} \;+\; \sigma(T) r \;+\; C,
\label{eq:cornell}
\end{equation}
is also shown using continuum parameters $\alpha_{\rm c}=\pi/12$
\cite{Luscher:1980ac} and $\sqrt{\sigma}=0.445$ GeV as used in
\cite{Satz:2005hx} (with $C$ adjusted to overlie our data). It was
shown in \cite{Satz:2005hx} that these parameters reproduce the
properties of the lowest lying states in both charmonium and
bottomonium very accurately. Since our ``zero'' temperature
(i.e. $T=0.24T_{\rm C}$) data follow this established Cornell
potential extremely well, they provide strong evidence that our method
is extracting the correct physics.


\begin{figure}
\centering
\includegraphics[scale=0.7, trim = 0 420 0 0, clip=true]{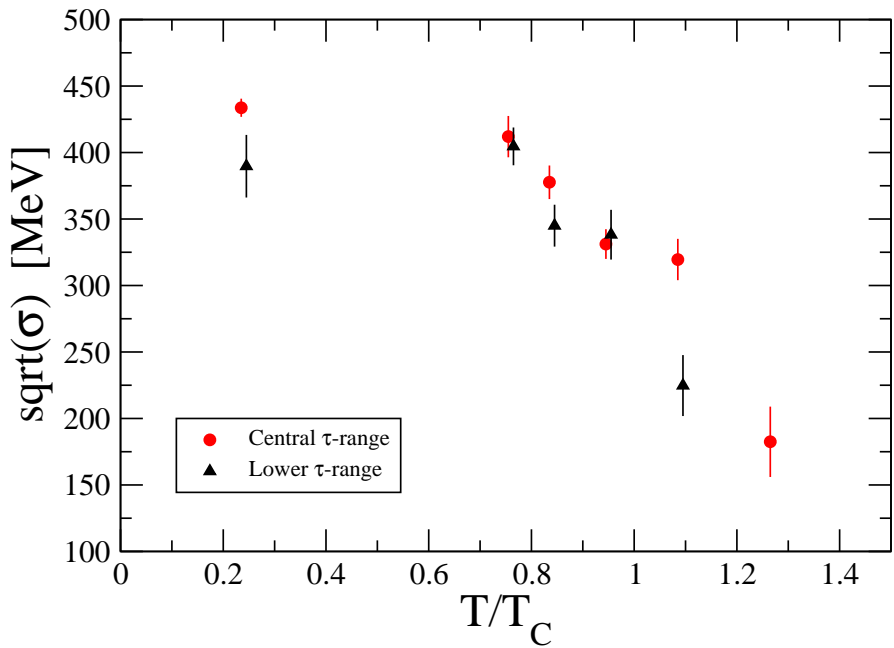}
\caption{The string tension, $\sigma$, as a function of temperature
  obtained by fitting the potential data to the Cornell potential
  \eqref{eq:cornell}.  Two different $\tau-$ranges ``best'' and
  ``lower'' were also used so that temperature effects can be
  uncovered, see Table \ref{tab:nf2plus1cratau}. The points are
  shifted slightly horizontally for clarity.}
\label{fig:string-tension}
\end{figure}

The temperature dependence of the central potential is further studied
by fitting the $V_{\rm C}$ data to the Cornell function
\eqref{eq:cornell} for both the ``best and ``lower'' $\tau$ ranges,
see Table \ref{tab:nf2plus1cratau}.  We restrict the fit to
separations in the range $2a_s \le r \le 9a_s$ due to the systematics
at very small and large separations discussed in section
\ref{sec:central}.  The parameters from these fits are shown in Table
\ref{tab:stringtensions} and the resultant string tensions are plotted
in Figure \ref{fig:string-tension}.  We also show the fits in Figure
\ref{fig:Vc_with_Cornell} for the coldest and hottest temperatures.

At face value, there is a clear temperature dependence in the string
tension as displayed in Figure \ref{fig:string-tension}. However,
using the strict criteria outlined in section \ref{sec:central}, the
neighbouring temperature's ``best'' and ``lower'' string tensions
overlap within errors, and so higher statistics are required to
properly decouple this thermal effect from the systematics in this
quantity.

A calculation of the string tension at zero temperature was performed
in \cite{Kawanai:2011jt} with nearly physical light dynamical quarks
using the {\sc hal qcd} method, where they found $\sigma = 394(7)$
MeV.  In studies using the static quark potential, the string tension
displays clear temperature effects well below $T_C$, see
e.g. \cite{Cardoso:2011hh} (in contrast to Figure
\ref{fig:string-tension}).  However \cite{Cardoso:2011hh} study pure
gauge $SU(3)$ where the confining string tension is essentially an
order parameter which is therefore zero for $T>T_{\rm C}$.

In \cite{Karsch:1987pv}, an alternative to the Cornell potential
\eqref{eq:cornell} was proposed for finite temperature where there is
colour-Debye screening of the colour sources. This has the form,
\begin{equation}
V(r,T) = - \frac{\alpha_{\rm s}(T)}{r}e^{-m_D(T)r}
\;+\; \frac{\sigma(T=0)}{m_D(T)} \left(1 - e^{-m_D(T)r}\right)
\;+\; C,
\label{eq:screened}
\end{equation}
where $m_D(T)$ is the Debye screening mass. This functional form has
the feature that $V(r,T)$ remains finite as $r \rightarrow \infty$.

We have performed fits of the central potential for each temperature
using \eqref{eq:screened} with three fitting parameters, $\alpha, m_D$
and $C$. We fixed $\sigma$ to its ``zero'' temperature value (obtained
with our $T=0.24T_{\rm C}$ ensemble) of 434 MeV, see Table
\ref{tab:stringtensions}.  As the temperature increases, the screened
form, \eqref{eq:screened}, fits the data better than the Cornell form,
\eqref{eq:cornell}. This can be seen in Figure
\ref{fig:Vc_with_Cornell} where both the screened and Cornell fit for
the hottest temperature are shown.  For the coldest temperature, the
screened and Cornell fits are indistinguishable.

\begin{figure}
\centering
\includegraphics[scale=0.7, trim = 0 420 0 0, clip=true]{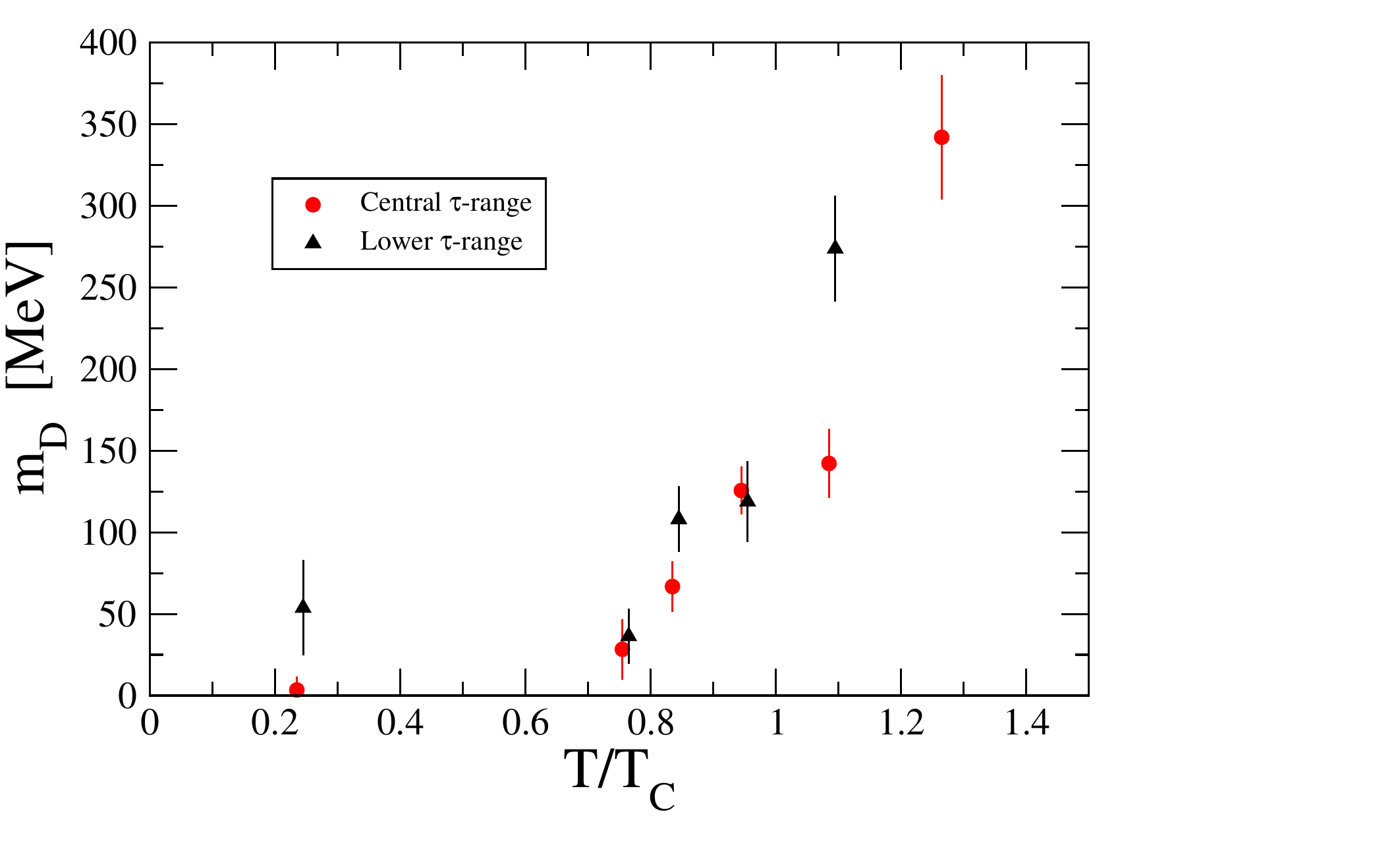}
\caption{The Debye mass, $m_D$, as a function of temperature
  obtained by fitting the potential data to the screened potential
  \eqref{eq:screened}.  Two different $\tau-$ranges ``best'' and
  ``lower'' were also used so that temperature effects can be
  uncovered, see Table \ref{tab:nf2plus1cratau}. The points are
  shifted slightly horizontally for clarity.}
\label{fig:m_D}
\end{figure}

The results of these fits are shown in Table \ref{tab:stringtensions}
and the resultant Debye masses are plotted in Figure \ref{fig:m_D}.
At low temperatures, $m_D\sim 0$ and it then has a rapid increase
around $T_{\rm C}$, in agreement with expectations. In
\cite{Digal:2005ht}, the Debye mass was calculated from a two-flavour
lattice calculation of the static quark-antiquark free energy. They
found very similar behaviour to Figure \ref{fig:m_D}, with $m_D$
increasing rapidly around $T_{\rm C}$ to around 400 MeV at $1.2 T_{\rm
  C}$.


\begin{table}
\begin{center}
\begin{tabular}{cc|cc|cc}
\hline
&&&&&\\
$T/T_C$ & $\tau$-range 
 & \multicolumn{2}{c}{Cornell Fit \eqref{eq:cornell}}
 & \multicolumn{2}{|c}{Screened Fit \eqref{eq:screened}} \\
&&&&&\\
 && $\sqrt{\sigma}$ &  $\alpha_{\rm c}$ &
    $m_D$ & $\alpha_{\rm s}$ \\
&& [MeV] && [MeV] & \\
&&&&&\\
\hline
0.24     & 30-63   & 434(7)  & 0.31(2) &   3(8)  & 0.311(9) \\
0.24     & 15-19   & 390(20) & 0.39(5) &  50(30) & 0.33(3)  \\
\vspace*{-3mm} &&&&& \\
0.76     & 15-19   & 410(20) & 0.36(3) &  30(20) & 0.34(2)  \\
0.76     & 12-17   & 405(14) & 0.41(3) &  40(20) & 0.38(2)  \\
\vspace*{-3mm} &&&&& \\
0.84     & 12-17   & 378(13) & 0.41(3) &  70(20) & 0.35(2)  \\
0.84     & 11-15   & 340(20) & 0.46(3) & 110(20) & 0.36(2)  \\
\vspace*{-3mm} &&&&& \\
0.95     & 11-15   & 331(11) & 0.48(2) & 130(20) & 0.37(2)  \\
0.95     & 11-13   & 340(20) & 0.49(3) & 120(30) & 0.39(2)  \\
\vspace*{-3mm} &&&&& \\
1.09     & 11-13   & 320(20) & 0.43(3) & 140(20) & 0.31(2)  \\
1.09     &  9-11   & 220(20) & 0.58(3) & 270(30) & 0.42(3)  \\
\vspace*{-3mm} &&&&& \\
1.27     &  9-11   & 180(30) & 0.53(3) & 340(40) & 0.37(3) \\
\hline
\end{tabular}
\caption{The results of fitting the central potentials to both the
  Cornell form, \eqref{eq:cornell}, and the Debye-screened form,
  \eqref{eq:screened}, as described in the text.  Note that, following
  the procedure outlined in section \ref{sec:central}, two
  $\tau-$ranges were chosen to allow a direct comparison between
  neighbouring temperatures.}
\label{tab:stringtensions}
\end{center}
\end{table}



\subsection{Spin Dependent Potential}
\label{sec:spin}

\begin{figure}
\centering
\includegraphics[scale=0.7, trim = 0 420 0 0, clip=true]{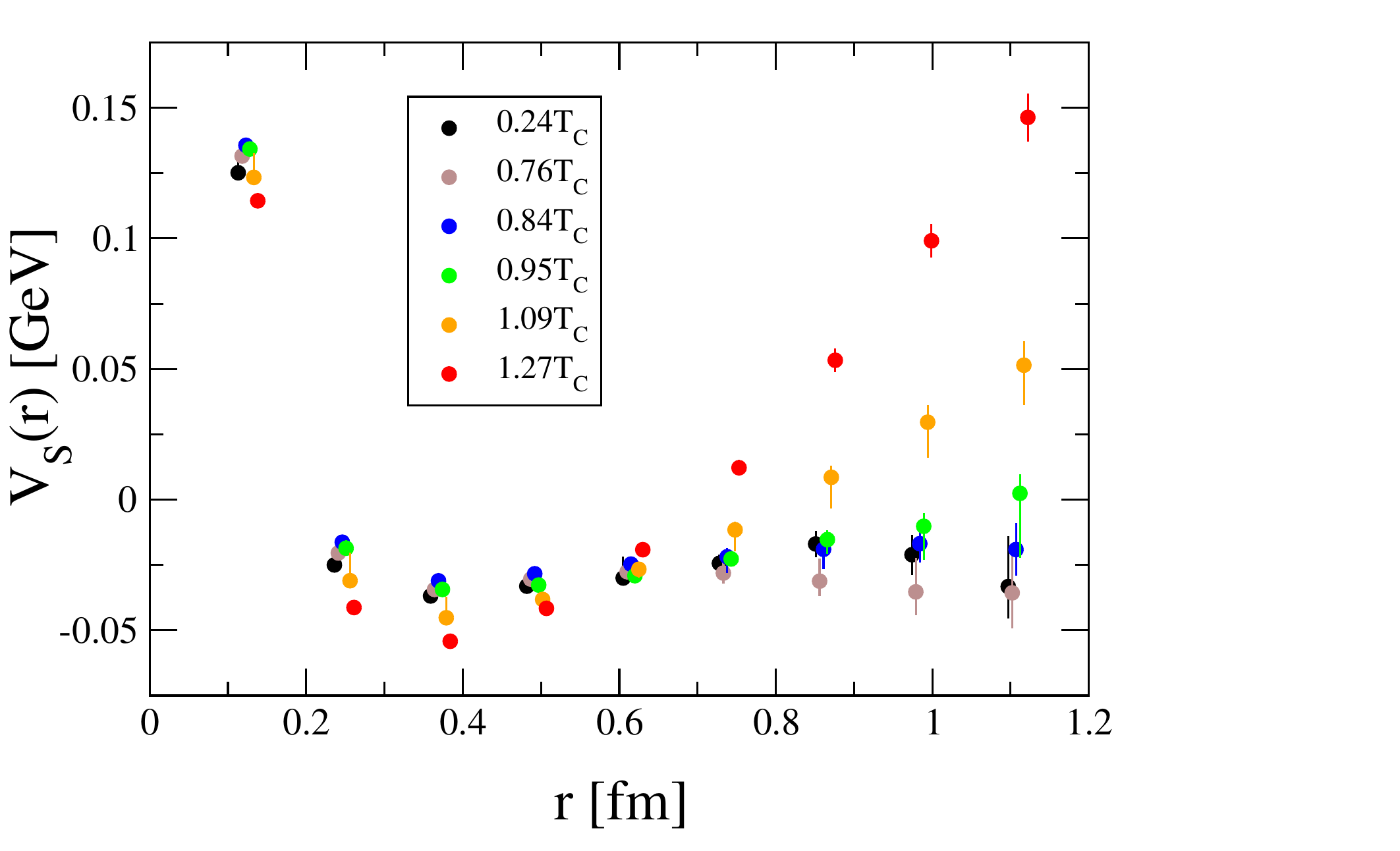}
\caption{The spin-dependent
  potential with combined statistical and systematic errors for
  various temperatures. The points are shifted slightly horizontally
  for clarity.}
\label{fig:Nf2plus1_Vs_aM_all_Ntau}
\end{figure}

We also obtained the ($\tau-$dependent) spin-dependent potentials for
the different temperatures by combining the pseudoscalar and vector
time-slice potentials from each ensemble according to
\eqref{eq:V_CS}. The $\tau-$dependence was removed using the same
procedure as in the central potential case to obtain Figure
\ref{fig:Nf2plus1_Vs_aM_all_Ntau}.

Taken at face value, we see a strongly repulsive core, but this has to
be qualified by the systematics in the $r/a_s=1$ data point
as discussed in section \ref{sec:central}.
However, since the spin-dependent potential is the {\em difference},
${V_V-V_{PS}}$, the systematics at $r=a_s$ may cancel to some
extent. Also note that
\cite{Kawanai:2011xb,Kawanai:2011fh,Kawanai:2011jt,Kawanai:2013aca}
have found a repulsive core for this potential, but with a quenched
calculation. Modelling the interaction via one-gluon exchange, the
spin-dependent potential is a $\delta-$function at the origin, so
given the body of lattice results including this work, a finite-width,
repulsive core appears to be the correct, non-perturbative result.
Note also the work of \cite{Laschka:2012cf} where a finite-width
repulsive potential was obtained by including the running of the
coupling in one-gluon exchange.

For confined temperatures, $T\ltap T_C$, the potential is clearly flat
for moderate to large distances with no significant temperature
dependence, see Figure \ref{fig:Nf2plus1_Vs_aM_all_Ntau}.  In our
calculation, as in \cite{Kawanai:2011xb,Kawanai:2011fh} the asymptotic
value for this confined phase potential is negative. However, when the
spin-dependent potential is used to define dynamically the reduced
mass, $\mu$, (see \eqref{eq:hal_schro}) this potential tends to zero
at large distances \cite{Kawanai:2011jt,Kawanai:2013aca} by definition.

There is a clear temperature effect once the deconfined phase is
reached with a distinct minimum at intermediate distances $r\sim 0.4$ fm
and significantly larger potential values at large distances $r\gtap
0.7$ fm compared to the same potential in the confined phase.


\subsection{Comparison with other methods}

\begin{figure}
\centering
\includegraphics[scale=0.7, trim = 0 420 0 0, clip=true]{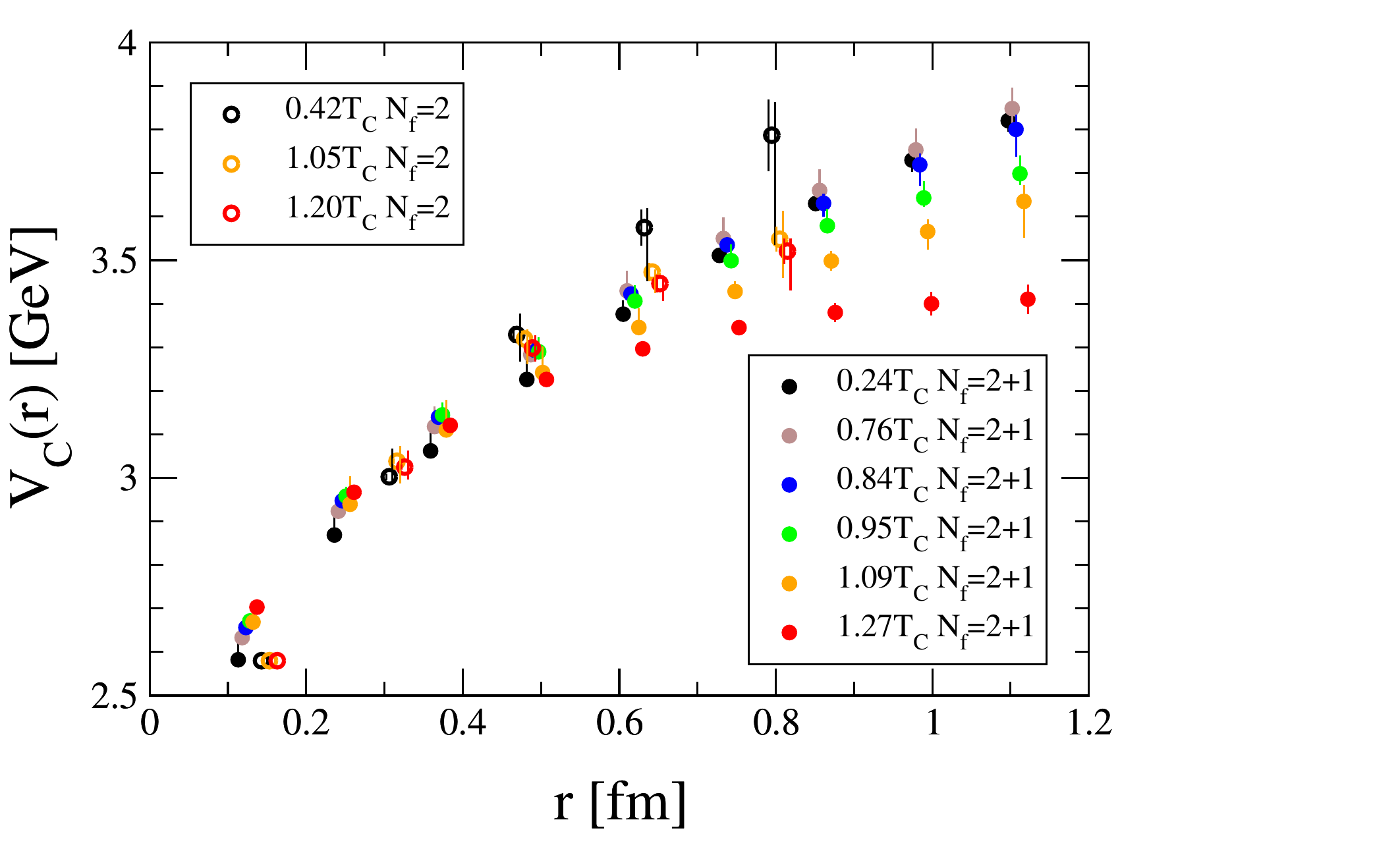}
\caption{A comparison
  of the central potentials obtained from this ${N_f=2+1}$ work and
  our earlier ${N_f=2}$ simulations \cite{Evans:2013yva}.  The
  ${N_f=2}$ data have two error bars, statistical (left) and
  systematic (right), and are shifted vertically so that the potential
  at the first separation agrees with the $N_f=2+1$, $0.24T_C$
  values. All points have been shifted slightly horizontally for
  clarity.}
\label{fig:Nf2plus1_vs_Nf2_td}
\end{figure}

In Figure \ref{fig:Nf2plus1_vs_Nf2_td} our central potentials from
this work (i.e. using our ${N_f=2+1}$ 2nd generation ensembles) are
compared with those obtained from our earlier, ${N_f=2}$ 1st
generation simulations \cite{Evans:2013yva}. The $N_f=2$ potentials
were also obtained with the {\sc hal qcd} time-dependent method, and
have been shifted vertically in Figure \ref{fig:Nf2plus1_vs_Nf2_td} so
that their $r/a_s=1$ data points coincide with that of the $0.24T_{\rm
  C}$ potential. It is encouraging that the potential data points
interpolate each other at small $r$, especially given that the lattice
parameters and actions used in each simulation are quite
different. For a given temperature the ${N_f=2+1}$ central potentials
are flatter at large $r$ than those from the ${N_f=2}$
simulation. This could be due to the inclusion of an extra sea quark
that has the ability to screen the strong force between quarks, but
further studies would be required to confirm this.


\begin{figure}
\centering
\includegraphics[scale=0.7, trim = 0 420 0 0, clip=true]
{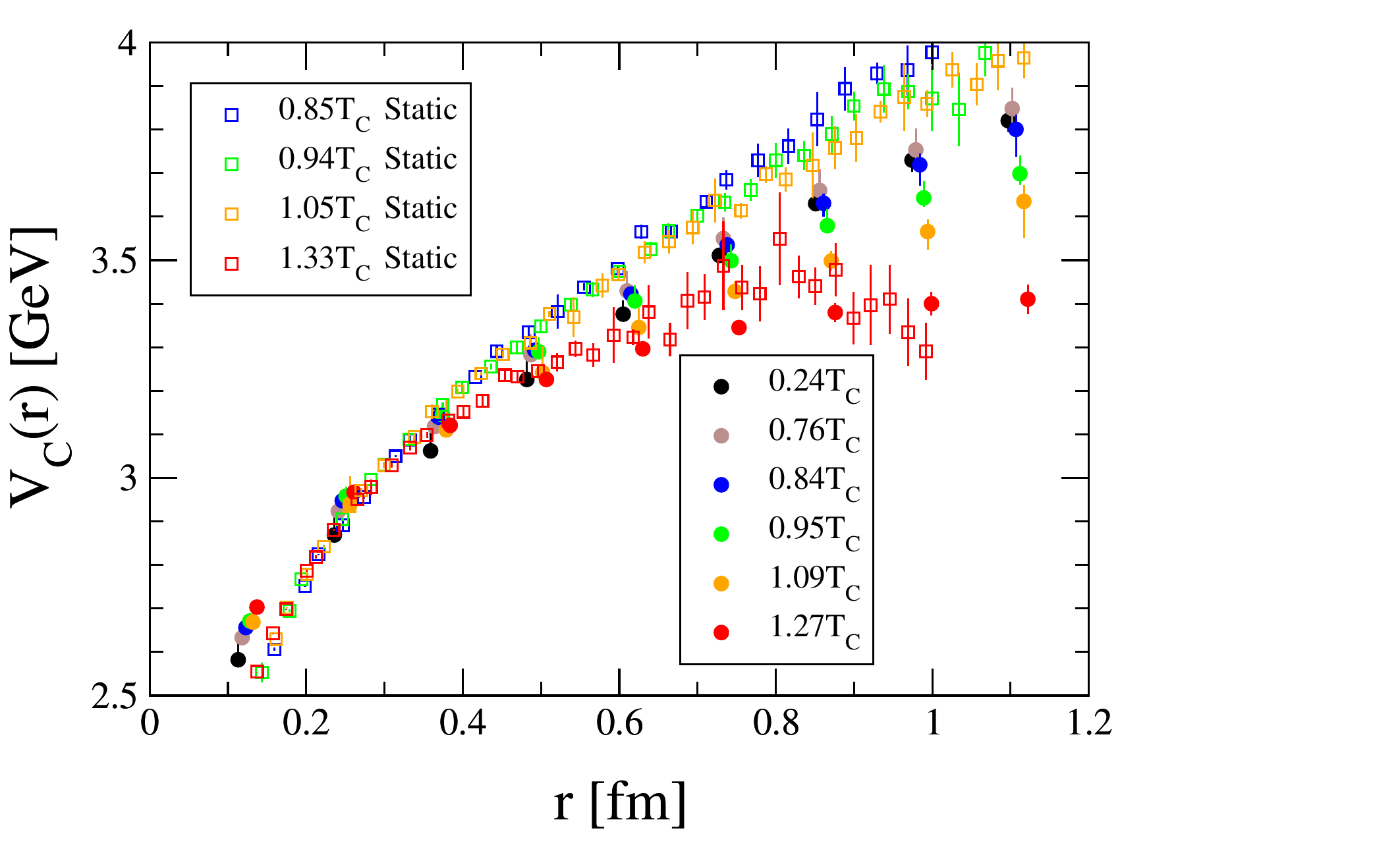}
\caption{A comparison of the central potential obtained from this work
  with the static quark potential for comparable temperatures obtained
  in \cite{Burnier:2014ssa}. The static potential data have been
  shifted vertically for the sake of comparison.}
\label{fig:static}
\end{figure}

In Figure \ref{fig:static} we compare the central potentials with the
static quark potentials calculated from the Wilson lines
\cite{Burnier:2014ssa} which were also obtained with 2+1 flavours.  The
static quark potential curves in Figure \ref{fig:static} are shown at
the temperatures closest to those in this work, and have been shifted
vertically so that their form can be compared to our
result.\footnote{This is justified since in the static limit the quark
  mass, which sets the overall scale, has been removed from the free
  energy calculation.} While the higher temperature results agree
fairly well between the two approaches, the lower temperature static
data are steeper than our results.  Further study would be required to
determine if this difference is due to \cite{Burnier:2014ssa} using
the infinite quark mass approximation rather than the physical charm
quark, or to other systematic differences between the two approaches.


\section{Conclusions}
\label{sec:conclusion}

There is a significant body of theoretical work studying the
interquark potential at non-zero temperature using model, perturbative
and lattice (non-perturbative) approaches. However, until
\cite{Iida:2011za,Evans:2013yva}, these lattice studies all used the
static quark limit. This work improves upon these static calculations
by considering quarks with finite mass, and thus represents a
first-principles calculation of the charmonium potential of QCD at
non-zero temperature. The method we used is based on the {\sc hal qcd}
time-dependent approach which obtains the potential directly from
local-extended correlators.

We do not observe any significant temperature dependence of the
central potential below $T_C$, while there is a significant flattening
above $T_C$, consistent with the expectation that the potential
becomes deconfining. The string tension is calculated and we find a
slower variation of this quantity with temperature than that found
using the static quark approximation. Using the Debye-screened form
for the potential (which fits our data better than the Cornell form),
we determine the Debye mass which is found to be very small at low
temperatures and then increase rapidly around $T_{\rm C}$.  This is,
as far as we know, the first non-perturbative calculation of the Debye
mass in charmonium. In the case of the spin-dependent potential, we
similarly find no thermal modification for $T\ltap T_C$, but a clear
variation with temperature at large distances in the deconfined phase
and evidence for a repulsive core.

This work improves upon our earlier work \cite{Evans:2013yva,
  Allton:2013wza} by including a dynamical strange quark, and using
lattices which are finer, with a larger volume, and have lighter, more
physical $u,d$ quarks.



\acknowledgments

We acknowledge the support and infrastructure provided by the Irish
Centre for High-End Computing, HPC Wales, the UK DiRAC Facility
jointly funded by STFC, the Large Facilities Capital Fund of BIS and
Swansea University, and the PRACE grants 2011040469 and
2012061129. The calculations were carried out using the Chroma
software suite \cite{Edwards:2004sx}. We thank Renwick (Jamie)
Hudspith for providing rapid gauge-fixing software, and also Balint
Jo\'o and Robert Edwards for hosting PWME's academic visit to
Jefferson Lab funded by the Welsh Livery Guild Travel Scholarship, and
Gert Aarts, Sinya Aoki and Paul Rakow for useful conversations.
PWME and CA are supported by the STFC, and PWME acknowledges the
support of the ERC.



\appendix
\section{Appendix: Momentum Space Propagators}
\label{sec:momspace}

Local-extended correlators can be obtained more efficiently by working
with the quark propagators in momentum rather than coordinate space
\cite{privsin}. While this method (which is not our own) has been
known for some time and used in many papers which have calculated
potentials from the {\sc hal qcd} method and in studies of multi-baryon
states \cite{Doi:2012xd}, it does not appear to be in any publication,
so we outline it here for reference.

For a meson, the local-extended correlator, \eqref{eq:le_cf}, in the
gauge fixed case (with $U(x,x\!+\!\mathbf{r})$ set to unity) can be
written,
\begin{align}\nonumber
C_\Gamma(\mathbf{r},\tau)
&= -\sum_{\mathbf{x}} \Tr \left[
D^{-1}(\mathbf{x}\!+\!\mathbf{r},\tau\!:\!\mathbf{0},0)\Gamma
D^{-1}(\mathbf{0},0\!:\!\mathbf{x},\tau)\Gamma^\dagger
\right] \\
&= -\sum_{\mathbf{x}} \Tr \left[
D^{-1}(\mathbf{x}\!+\!\mathbf{r},\tau\!:\!\mathbf{0},0)\Gamma
\gamma_5 D^{-1}(\mathbf{x},\tau\!:\!\mathbf{0},0)\gamma_5 \Gamma^\dagger
\right]
\label{eq:old}
\end{align}
where $D^{-1}({\rm sink}\!:\!{\rm source})$ is the quark propagator.
This correlator can be written in terms of the Fourier transform of
the quark propagators,
\begin{equation}
D^{-1}(\mathbf{y},\tau\!:\!\mathbf{0},0)
= \frac{1}{V}\sum_{\mathbf{q}}\tilde{D}^{-1}(\mathbf{q})e^{i\mathbf{q}\cdot\mathbf{y}},
\end{equation}
giving
\begin{align}\nonumber
C_\Gamma(\mathbf{r},\tau) 
&=-\frac{1}{V}\sum_{\mathbf{p}}\Tr\left[
  \tilde{D}^{-1}(\mathbf{p}) \Gamma
  \gamma_5\tilde{D}^{-1}(-\mathbf{p})\gamma_5 \Gamma^\dagger
\right] e^{i\mathbf{p}\cdot\mathbf{r}} \\
&\equiv\frac{1}{V}\sum_{\mathbf{p}}\tilde{C}(\mathbf{p},\tau)e^{i\mathbf{p}\cdot\mathbf{r}},
\label{eq:Cfourier}
\end{align}
where $\tilde{C}(\mathbf{p},\tau)$ is the Fourier transform of the
correlator, $C_\Gamma(\mathbf{r},\tau)$.

This implies that once $\tilde{C}_\Gamma(\mathbf{p},\tau)$ is
obtained, then the desired correlator, $C_\Gamma(\mathbf{r},\tau)$,
can be simply computed for any value of $\mathbf{r}$ using the single
sum in \eqref{eq:Cfourier}. This is computationally more efficient
than having to perform the trace in \eqref{eq:old} for each
$\mathbf{x}$ value before then also performing the sum over
$\mathbf{x}$.


\clearpage
\bibliographystyle{JHEP}
\bibliography{./bibliography} 

\end{document}